\documentclass[smus]{snow2e_d}
\usepackage{epsfig}
\begin{document}
\hyphenation{counter-terms}

\begin{titlepage}

\headsep 2cm
\baselineskip 0.5cm

\evensidemargin 0.7cm

\begin{flushright}
{\large FERMILAB-CONF-97/ 018-T}
\end{flushright}

\vspace{1.5cm}

\begin{center}
{\Large \bf Summary of the Very Large Hadron Collider Physics and 
Detector Subgroup}\footnote{\large To appear in the {\it Proceedings of
the 1996 DPF/DPB Summer Study on New Directions for High-Energy Physics}
(Snowmass 96).}

\vspace{2cm}
{\Large D.~Denisov and S.~Keller} \\
\vspace{0.2cm}
{\large \it Fermilab, P.O.~Box 500, \\
Batavia, IL 60510, USA}
\vspace{6cm} 

{\Large \bf Abstract}\\
\vspace{0.5cm}
{\large 
We summarize the activity of the Very Large Hadron Collider 
Physics and Detector subgroup during Snowmass 96.
}
\end{center}

\end{titlepage}

\title{Summary of the Very Large Hadron Collider Physics and Detector Subgroup
}

\author{D.~Denisov and S.~Keller \\ {\it Fermilab, P.O. Box 500,
Batavia, IL 60510, USA}}
 
\maketitle

\thispagestyle{empty}\pagestyle{empty}

\begin{abstract} 
We summarize the activity of the Very Large Hadron Collider 
Physics and Detector subgroup during Snowmass 96.
Members of the group: M. Albrow, R. Diebold, S. Feher, L. Jones, R. Harris, D. Hedin, 
W. Kilgore, J. Lykken, F. Olness, T. Rizzo, V. Sirotenko, and J. Womersley.
\end{abstract}

\section{Introduction}
Considering the long lead time for accelerator projects it is important
for us to investigate possible options for colliders beyond the LHC.  
This subgroup was motivated by the accelerator work~\cite{PIPE} that 
has been started on new technologies for a post-LHC 
Very Large Hadron Collider (VLHC)~\cite{RLHC}.

The goal of this subgroup was to start the discussion 
on physics and detector issues associated with 
a VLHC with an energy in the center of mass
of the order of 100 to 200 TeV and a luminosity up to $10^{34-35} cm^{-2} s^{-1}$.  
Obviously, physics and detector issues, along with accelerator 
technology and budget constraint, must guide us to select appropriate and realistic 
energy and luminosity for such a machine. 
Defining the physics goals of a post-LHC accelerator is not trivial.  
As is well known, the last largely unexplored sector of the Standard Model (SM), 
the Higgs sector, will be investigated over the next decade by the Tevatron, LEP, and LHC.
It is therefore likely that any post-LHC 
machine will be built to explore Physics beyond the SM.
At this point in time, we do not have any experimental evidence for the physics 
beyond the SM, and it is therefore difficult to make the case for a specific accelerator
beyond the LHC.  Our goal is to 
make the case for R\&D on accelerator and detector technologies 
that would allow us to build a VLHC with a lower cost than current methods.   

We have to investigate different models/scenarios of physics
beyond the SM and understand their implications for a VLHC.  
It is important to provide luminosity requirement versus the 
energy of the machine for fixed physics goal(s). 
It seems that to be successfull any new accelerator will need to probe physics 
at a scale at least an order of magnitude larger than the LHC.  

We had a few meetings before Snowmass where we defined
possible topics to work on during Snowmass.  
For the physics issues we used the EHLQ~\cite{EHLQ} paper as a guide.  
As the physics that will be probed by this machine is not yet known,
a multipurpose detector was discussed.

\section{Summary of work done at Snowmass}
In this section, we briefly summarize the work that was done during the 
workshop by the different members of the group.

F.~Olness: ``Can we use the current distribution function sets?''.
The upper limit for the factorization scale, $Q$, of the parametrized version of 
CTEQ3M is 10TeV.  The set still can return values 
at higher value of $Q$.
The Bjorken-$x$ and Q variation behave as expected, with the $q\bar{q}$, $qg$, and 
$gg$ fluxes dominating at 
high, intermediate, and low Bjorken-$x$, respectively.  
Only the valence $d$-quark distribution seem to behave abnormaly: it starts rising at 
10TeV for $x\sim .5$.   The other distributions appear to be fine in that range.
The momentum sum rule is a good check that can be performed.  
It is off by about 5\% at 10TeV and by about 15\% at 100TeV.
In conclusion, for the current level of accuracy needed, 
CTEQ3M seems to be adequate.  
It would be convenient if, in the future, CTEQ and other groups 
would generate sets that can be used up to 100 TeV.

S. Keller: ``$W'$ production''. 
This simple example can be used to illustrate the basic features of going 
to higher energy.  The same couplings as for the $W$ are used, only the 
mass is changed.  Using the narrow width approximation one can 
easily derive the following expression for the cross section:
\begin{eqnarray}
\sigma (E_{cm}) \sim {\frac{1}{E_{cm}^2}}
 \sum_{a,b} 
\int_{\frac{M_{W'}^2}{E_{cm}^2}}^1
\frac{dx_1}{x_1} F_a(x_1,Q) F_b(x_2,Q),
\label{eq:sigma}
\end{eqnarray}
where $E_{cm}$ is the center of mass energy of the collider, 
$F_a$ is the parton distribution function of parton $a$ inside of the proton (or anti-proton)
and $x_i$ the Bjorken-$x$ of the partons.
From this expression the scaling rule that is often used can be explained:
an increase of the energy by a factor $n$ require an increase in the
luminosity by a factor of $n^2$ ( due to the factor $1/E_{cm}^2$) in order to maintain a
constant number of events.   The luminosity must be increase by a factor of 
about 200 when going from the LHC energy of 14 TeV to a VLHC at 200 TeV!  
However, this is only true if the rest of the expression in Eq.~\ref{eq:sigma} 
is kept constant, in other word if $M_{W'}/E_{cm}$ is kept constant and the 
$Q$ evolution of the distribution function is neglected.  
Keeping the ratio $M_{W'}/E_{cm}$ constant is synonymous of maximizing the machine potential:
with a luminosity 200 times bigger than at the LHC, a VLHC at 200 TeV would also investigate
a mass of the $W'$ that is about 14 times bigger than at 
the LHC.  On the other hand, if the goal is to compare the physics potential of different 
machines then the physics goal ($M_{W'}$ in this case) should be fixed.  Increasing
the energy then reduces the value of Bjorken-$x$ probed.
This dramatically increases the cross section because of the increase in the distribution 
function at small $x$. 
For example, for $M_{W'}=25TeV$, with each doubling of the energy from 
50 to 200 TeV, there is roughly an increase in the cross section by a factor of 10 and 
therefore a reduction by a factor of 10 in the integrated luminosity
needed to discover the $W'$ at that fixed mass.  
For this particular process, 
a VLHC at $E_{cm} =200 TeV$ and a luminosity ${\cal L} = 10^{34} cm^{-2} s^{-1}$ 
is equivalent to a VLHC at  $E_{cm} =100 TeV$ and ${\cal L} = 10^{35} cm^{-2} s^{-1}$.
Obviously, even without increasing the luminosity compared to the LHC, 
a VLHC with a center of mass energy in the 100 to 200 TeV range 
will probe physics at a scale higher than at the LHC.

$W'$ production is also a usefull example to compare the $pp$ and $p\bar{p}$ options.  
$W'$ is produced through a $q\bar{q}$ pair and it is therefore expected that 
for large values of $M_{W'}$ the $p\bar{p}$ will have the advantage over the 
$pp$ option, because the valence-valence flux is bigger than the valence-sea flux at high
Bjorken-$x$.   The cross section is at most 3--4 times larger in $p\bar{p}$ than in $pp$
at high Bjorken-x, for $M_{W'} \sim E_{cm}/2.$.  The luminosity for the $p\bar{p}$ option
can therefore be decreased at most by the same factor, but this advantage should be gauged 
against the potential loss for other processes.
    
J.~Lykken:  ``Supersymmetry and the VLHC''.
There is a consensus opinion that if weak-scale SUSY exists, the LHC
will see it and that the LHC and a 1-1.5 TeV NLC are sufficient to do a
good job on s-particle spectroscopy.
On the other hand, any SUSY discovery immediately implies the existence of
at least two new physics scales beyond the weak scale: the dynamical SUSY
breaking scale (hidden sector) and the messenger scale.
In SUGRA models these scales are close to the Planck scale and not
accessible, but in gauge-mediated models the hidden sector scale can be
in the range $10^2-10^4$ TeV and the messenger scale is likely in the
range $10-100$ TeV if one assumes that the gluino mass is in the
100-1000 GeV range. If SUSY is discovered, we will know if it is
gauge-mediated versus gravity-mediated from distinctive signatures
like $\chi_1^0 \rightarrow \gamma +gravitino$,
which can be seen at LEPII or the Tevatron.
Measuring the $\chi_1^0$ lifetime from displaced vertices at LHC can
tell us the scale of the messenger.
If this scenario is correct, we may get a ``no-loose'' theorem for a VLHC.

W.~Kilgore (with M.~Peskin):  ``Multiple Production of $W$'s and
$Z$'s''.  In the event of a strongly coupled electroweak symmetry
breaking sector, multiple $W$ and $Z$ production could be large enough
to dominate the $W+n$ jets cross section at large $n$ at a VLHC.  This
would be similar to multiple pion production in QCD but would involve
the ``pions'' ({\it i.e.\/} longitudinal $W$'s and $Z$'s) of the
strongly interacting symmetry breaking sector and would help to
characterize that sector.

T.~G.~Rizzo~\cite{RIZZO}: ``Searches for Scalar and Vector Leptoquarks, 
Searches for New Gauge 
Bosons, and Constraints on $q\bar{q} \gamma \gamma$ Contact Interactions''.  
1) Both scalar and vector leptoquarks should easily be discovered at the LHC if their masses
are not in excess of the 1--2 TeV range.  A VLHC ($pp$ collider) with an energy of 200TeV in 
the center of mass and a integrated luminosity of $10^3fb^{-1}$ would increase the 
search significantly 
by an order of magnitude to the 10--20 TeV range.  
2) The search reaches for new gauge bosons are summarized for 
a variete of extended gauge models.  Here too a VLHC would increase the mass search reach by 
an order of magnitude from the $\sim$ 5TeV range at the LHC to the $\sim$ 50 TeV range.     
3) High $p_t$ diphoton events with large invariant masses put constraints on 
flavor-independent $q\bar{q} \gamma \gamma$ contact interactions.  Constraints on the 
corresponding $e^+e^- \gamma \gamma$ contact interaction already exist from LEP.  
Constraints on the scale associated with these contact interactions are improved by a factor
of about 8 and 5 compared to the bounds provided by the LHC for the case of constructive and
destructive interference, respectively.   In the three cases studied, the signal is initiated
by $q\bar{q}$ such that there is an improvement in the mass or scale reach 
of about $20-40\%$ when switching to a $p \bar{p}$ collider. 

R.~M.~Harris~\cite{HARRIS}: ``Discovery Mass Reach for Excited Quarks at Hadron Colliders''.
If quarks are composite particles then excited states are expected.  
For an integrated luminosity of $100fb^{-1}$ ($10^{34}cm^{-2}s^{-1}$ for a year)
the mass reach is increased from 6TeV at the LHC to 18 (31, 52) at the VLHC
for an energy in the center of mass of 50TeV (100TeV, 200 TeV).
Suppose that the LHC sees a classic signal of new physics: an excess of 
high energy transverse energy jets (assuming that we can rule out the excess
as due to the parton distribution functions).   This would be strong evidence
of new physics, but the nature of this new physics would not be that clear.
If the source of new physics is compositeness, we would then expect
to see excited quark.  Let's assume that we expect an excited quark at a mass
around 25TeV.  We would then need to decide wich machine to build to 
find that excited quark.  Although a 50TeV machine would require a 
luminosity of about $10^4fb^{-1}$, a 100 TeV and 200 TeV would only require
about $10fb^{-1}$ and $1fb^{-1}$.  The current wisdom that a factor of 2 
in energy is worth a factor of 10 in luminosity is valid between the two
higher energy machine.  The factor is much bigger between the lowest and 
the two highest.  

D.~Hedin: ``Thoughts on Designing Detectors for the VLHC''. 
Simple scaling rules require a general purpose detector which is larger 
than LHC detectors.  However, considering that the cost of a detector 
is roughly proportional to its (size)$^{3}$ it is difficult to imagine 
how detectors larger than CMS or ATLAS can be built. 
However, large portion of the project cost is actually hidden within
physicists and engineers salaries. In order to minimize these costs we  
should not start from scratch, but consider to recycle the detector(s) with appropriate 
changes. If we consider CMS as an example, elements such 
as the magnet, muon iron, muon chambers and maybe even the calorimeters can be 
used at a VLHC, while DAQ, electronics, inner
tracking and trigger systems would have to be replaced. 

V.~Sirotenko: ``QCD jets at a VLHC''. 
As usual, soft processes will be a background to many processes of
interest.  Due to the large cross section, these soft interactions
determine the detector environment: occupancy, radiation doses, etc. 
PYTHIA 5.7 was used in order to simulate proton-proton collisions with 
a center of mass energy of 200TeV with cut parameter $P_t(min) > 25 GeV$ .  
The following results have been obtained:

a) charged particle multiplicity: 
$170$ for all $|\eta|$ and $50$ for $|\eta| < 3$ per event;

b) the $P_t$ spectra of charged particles is as expected very soft, the 
average value is around 0.8 GeV;

c) total energy flow for $|\eta| < 3$ is around 250GeV per event. 
In the central region energy flow in a $0.1 \times 0.1$ tower is 0.03GeV per
event.  Energy fluxes vs detector rapidity were simulated. There are concerns about
the accuracy of PYTHIA at such high center of mass energy and about the sensitivity
of the results on the choice of the $P_t(min)$ cut parameter.

J.~Womersley: ``Physics and detector issues for an O(100TeV) pp collider''. 
If we assume that a luminosity of 
${\cal L}=10^{36}cm^{-2}s^{-1}$ will be required for the physics,
and a bunch spacing time of 20~ns, then we will have 2000 interactions per
crossing or around $10^{5}$ charged tracks in $|\eta|$$<$2.5. Therefore
of order $10^7$ tracking elements would be required
to obtain an occupancy of a few percent; at a 
radius of 1~m, this would mean 1~mm$^2$ pixels, or 100~$\mu$m$^2$ at
10~cm radius. For 10 tracking
layers the total channel count would be $10^{8}$
and so even with a few \$$/$channel, the cost would
be very high. 
Electron identification in the presence of this large charged particle
flux will be very difficult:
the probability to find a ``track stub'' in front of an
electromagnetic calorimeter energy 
cluster is 20-100\%, so almost all $\pi^0$'s would look like electrons.
A central magnetic field is then an advantage to bend
slow particles and
provide an $E/p$ match. 
To determine muon momenta, the sagitta must be measured.  Since it
is propotional to $B*L^{2}$/p, to have a 
momentum resolution comparable with the LHC detectors at 10 times higher 
momentum one could improve
the coordinate precision by a factor of 10 (but the ATLAS/CMS goal is already
50~micron), increase the magnetic field
by a factor of 10 (to $\sim$40~T!) or increase the detector size by 
a factor of 3.  Some combination of the
above three possibilities is probably the best option.

D.~Denisov: ``Detector issues for 100-200TeV pp collider''.  
A high $P_T$ general purpose detector which can be built based on today technology 
is considered.  Heavy objects are produced almost at rest and the acceptance for their 
detection is proportional to the covered solid angle. 
The total cross section at 100TeV is about 150mb with an average number of charged particles 
per collision around $10^2$ ($|\eta|$$<$3).   There are three major elements 
in modern detectors: tracking, calorimetry and muon system.  
Tracking is required for electron identification.  
Occupancy and aging are two major problems for tracking. 
For ${\cal L}= 10^{35}cm^{-2}s^{-1}$ there would be of the order of $10^{4}$ tracks 
per crossing, assuming bunches are 20ns apart.  
About $10^{7}$ detector elements would be needed to keep occupancy low 
for a 10 layers tracking detector.
A magnetic field in the central region could bend soft particles away
from the central region and reduce occupancy
and radiation doses on the central tracker.  
Calorimetry resolution is just getting better with the increase in energy and
thickness has to increase only as the logarithm of the energy.  
The aging is the most serious problem for the calorimetry, along 
with underlying minimum bias events fluctuations.  High pressure gas ionization calorimetry
looks very promizing for VLHC applications, its radiation hardness is very good.  
Muon detection is not at all trivial.
Utilization of muon radiation at such high energy may be a better choice than the standard
saggita measurement.  Typical parameters for the interaction region should be in the following range: 
10 micron perpendicular to the beam axis, 1 m along the beam,  
and 10-20 ns interval between beam crossings.

R.~Diebold~\cite{DIEBOLD}: ``Physics per buck''.  A simple model to
characterize the cost-benefit ratio, the ``physics per buck'', for
the energy of future accelerator is described.  The new physics
capability is assumed to be proportional to the logarithm of the
ratio of the beam energies for the new and old facilities, whereas
the cost is assumed to have a fixed component plus one that is
linear with the energy of the beam.  The production of heavy
``$W'$'' (the same $W'$ model already mentioned in this section) is
used as a benchmark for the physics goal.  The optimization model is
insensitive to details and shows a broad maximum as a function of
accelerator energy, with a slow dependence on the ratio of fixed to
linear costs for the facility.  The model supports the common wisdom
and past choices of energy increases by a factor of roughly 3 to 15.
Different aspects of the approval process for a machine beyond the
LHC are also discussed.  The need to have a successfull LHC program
is stressed, along with the need for internationalization, and the
careful control of costs.

M.~Albrow and L.~W.~Jones~\cite{ALBROW}: ``Forward Physics with the
Pipetron''.
It is argued that provision should be made for a detector engineered 
to cover the complete final-particle phase space, 
i.e. to be sensitive to all rapidity including 
small angle, forward physics.
The agenda of such a detector would include a full range of physics
topics, mostly related to QCD. Rapidity gap physics should not be
compromised, including
hard single diffraction, and especially the study of
multi-TeV pomeron-pomeron interactions. 
The physics goals also relate to ultra high
energy cosmic ray observations.
Such a detector would require a long straight section ($\pm \sim 2$ km) 
in the accelerator design, and would 
work at a relative low luminosity ($\approx 10^{31}cm^{-2}s^{-1}$) with on average one
(or less) interaction 
per crossing.  The CDF and/or DO detectors could be recycled as central
detectors, and a series of forward detectors would cover large rapidities.
Such a detector should be included in the planning at an early stage
as it influences the lattice and tunnel design.

\section{Conclusions}

Obviously, the work that was done during Snowmass is not in any 
way completed, it is a start.  The following conclusions should be considered
within that context.
  
As the EHLQ~\cite{EHLQ} paper pointed out in its conclusions more than a decade
ago, there is no specific landmark in sight beyond the 1 TeV electroweak scale.
We still don't have any experimental evidence for physics beyond the SM, 
and therefore no clear-cut physics goal(s) for an accelerator beyond the LHC.

To be succesfull any post-LHC accelerator should explore scales at least 
an order of magnitude larger than the LHC.   For the examples studied 
during this workshop, a Very Large Hadron Collider with an energy in the 
center of mass of 200 TeV and a luminosity of $10^{34}cm^{-2} s^{-1}$
would achieve that goal, probing scales 5--10 times larger than the LHC
(14 TeV and $10^{34}cm^{-2} s^{-1}$).

For the collider energy range considered, 
a doubling of the energy is equivalent or better than an increase in 
the luminosity by a factor of 10.      
This seems to contradict one of EHLQ conclusions that stated the exact 
oposite for energy in the center of mass above 40 TeV.  
However, the scales considered in that work were an order of magnitude smaller 
than here.  As the scale considered increases, 
an increase in the energy becomes more important than an increase in 
luminosity and of course at the limit of that argument 
''no increase in luminosity can compensate for center of mass energy 
below the treshold for new phenomena'' (cf. EHLQ).  

The advantages of $p\bar{p}$ collisions over $pp$ collisions is limited to 
the production of heavy objects through a quark--antiquark initial state. 
The question here is wether or not the 
large luminosity required can be achieved for the $p\bar{p}$ option.

The physics studies can be separated in two groups.  
The first was centered around production models, comparing 
the reach of different colliders for the discovery of new particles.  
The second considered different scenarios 
of physics beyond the Standard Model that have a chance to reveal themselves
over the next decade (before a VLHC) and studied their implications for a VLHC.
For future studies, we think that 
it will be more interesting to concentrate on the second group.


VLHC detectors seem feasible using known
technologies.   There are many challenges, like keeping the total
number of tracks per crossing down to a manageable level,
measuring multi--TeV muons and finding materials that
can sustain the large radiation doses (possibly up to Trad in the forward 
region).  
Considering the projected cost of LHC detectors, it is clear that 
detectors should not be ignored in the overall cost of the 
project.  If the cost of the VLHC detectors stays at the same level than 
for the LHC, there is even an optimization to do: an increase in the accelerator energy
increases its cost, but allows to decrease 
the luminosity (for fixed physics goal(s)) and therefore more 
than likely to decrease the cost of the detectors.

The effort started at this workshop will continue.  Since the workshop we had
a few meetings~\cite{WEBPAGE} and a first workhop specifically
adressing the physics and detector issues associated with a Very Large 
Hadron Collider will be held at Fermilab on March 13-15 1997~\cite{WORKSHOP}. 
%
%


\begin{thebibliography}{20}
%
\bibitem{PIPE}
Mini-symposium ``New low-cost approaches to high energy hadron colliders 
at Fermilab'' at the May 3, 1996 APS Annual Meeting.  
G.~Dugan, P.~Limon, and M.~Syphers, ``Really Large Hadron Collider 
Working Group Summary'', these proceedings.  
E.~Malamud, `` New Technologies for a Future Superconducting 
Proton Collider at Fermilab'', these proceedings.
\bibitem{RLHC}
Some groups at this workshop
called this option the Really Large Hadron Collider (RLHC).  
\bibitem{EHLQ}
E.~Eichten, I.~Hinchliffe, K.~Lane, and C.~Quigg, 
Rev.~Mod.~Phys.~56:579-707,1984; Addendum-ibid.~58:1065-1073, 1986. 
\bibitem{RIZZO}
T.~Rizzo, these proceedings.
\bibitem{HARRIS}
R.~Harris, these proceedings and Fermilab-Conf-96/285-E, hep-ph/9609319.
\bibitem{DIEBOLD}
R.~Diebold, these proceedings.
\bibitem{ALBROW}
L.~W.~Jones, M.~Albrow, H.~R.~Gustafson, and C.~C.~Taylor, these proceedings.
\bibitem{WEBPAGE}
See our web page: http://www-theory.fnal.gov/vlhc/vlhc.html 
\bibitem{WORKSHOP}
For more information and to register see:
http://fnphyx-www.fnal.gov/conferences/vlhc/VLHCworkshop.html
or send e-mail to: denisovd@fnal.gov or keller@fnal.gov.
\end{thebibliography}
\end{document}